\documentclass[aps,prb,twocolumn,superscriptaddress,amsmath,amssymb]{revtex4-1}

\usepackage{graphicx}
\usepackage{color}

\newcommand{\La}[0]{La$_{2}$PtIn$_{8}$}
\newcommand{\Ce}[0]{Ce$_{2}$PtIn$_{8}$}
\renewcommand{\Pr}[0]{Pr$_{2}$PtIn$_{8}$}

\begin{document}
\title{Fermi-surface topology of the heavy-fermion system \Ce}

\author{J. Klotz}
\email{j.klotz@hzdr.de}
\affiliation{Hochfeld-Magnetlabor Dresden (HLD-EMFL), Helmholtz-Zentrum Dresden-Rossendorf, 01328 Dresden, Germany}
\affiliation{Institut f\"ur Festk\"orper- und Materialphysik, Technische Universit\"at Dresden, 01069 Dresden, Germany}

\author{K. G\"otze}
\altaffiliation[Present address: ]{Department of Physics, University of Warwick, Coventry, CV4 7AL, United Kingdom}
\affiliation{Hochfeld-Magnetlabor Dresden (HLD-EMFL), Helmholtz-Zentrum Dresden-Rossendorf, 01328 Dresden, Germany}
\affiliation{Institut f\"ur Festk\"orper- und Materialphysik, Technische Universit\"at Dresden, 01069 Dresden, Germany}

\author{E. L. Green}
\affiliation{Hochfeld-Magnetlabor Dresden (HLD-EMFL), Helmholtz-Zentrum Dresden-Rossendorf, 01328 Dresden, Germany}

\author{A. Demuer}
\affiliation{Laboratoire National des Champs Magn\'etiques Intenses (LNCMI-EMFL), CNRS, UGA, 38042 Grenoble, France}

\author{H. Shishido}
\author{T. Ishida}
\affiliation{Osaka Prefecture University, Department of Electronics, Mathematics, and Physics, Sakai, Osaka 5998531, Japan}

\author{H. Harima}
\affiliation{Graduate School of Science, Kobe University, Kobe 657-8501, Japan}

\author{J. Wosnitza}
\affiliation{Hochfeld-Magnetlabor Dresden (HLD-EMFL), Helmholtz-Zentrum Dresden-Rossendorf, 01328 Dresden, Germany}
\affiliation{Institut f\"ur Festk\"orper- und Materialphysik, Technische Universit\"at Dresden, 01069 Dresden, Germany}

\author{I. Sheikin}
\email{ilya.sheikin@lncmi.cnrs.fr}
\affiliation{Laboratoire National des Champs Magn\'etiques Intenses (LNCMI-EMFL), CNRS, UGA, 38042 Grenoble, France}

\date{\today}

\begin{abstract}

\Ce\ is a recently discovered heavy-fermion system structurally related to the well-studied superconductor CeCoIn$_{5}$. Here, we report on low-temperature de Haas-van Alphen-effect measurements in high magnetic fields in \Ce\ and \Pr. In addition, we performed band-structure calculations for localized and itinerant Ce-$4f$ electrons in \Ce. Comparison with the experimental data of \Ce\ and of the $4f$-localized \Pr\ suggests the itinerant character of the Ce-$4f$ electrons. This conclusion is further supported by the observation of effective masses in \Ce, which are strongly enhanced with up to 26 bare electron masses.

\end{abstract}

\maketitle

\section{Introduction}

Unconventional superconductivity in the vicinity of quantum-critical points (QCPs) is an intensely-studied phenomenon in Ce-based heavy-fermion (HF) systems. Among these materials, CeCoIn$_{5}$ exhibits the highest known superconducting transition temperature $T_{c} = 2.3$~K \cite{Petrovic_2001,Ikeda_2001}. Several theoretical works suggest that the presence of two-dimensional (2D) Fermi surface (FS) sheets enhances antiferromagnetic (AFM) fluctuations which are believed to be responsible for Cooper-pair formation in HF compounds \cite{Moriya_1990, Mathur_1998, Monthoux_1999, Monthoux_2003, Monthoux_2007, Shishido_2010}. Indeed, CeCoIn$_{5}$ possesses almost cylindrical FS sheets at low temperatures, where Ce-$4f$ electrons become itinerant \cite{Settai_2001, Hall_2001, Nomoto_2014}.

\begin{figure}
	\centering
		\includegraphics[width=0.9\columnwidth]{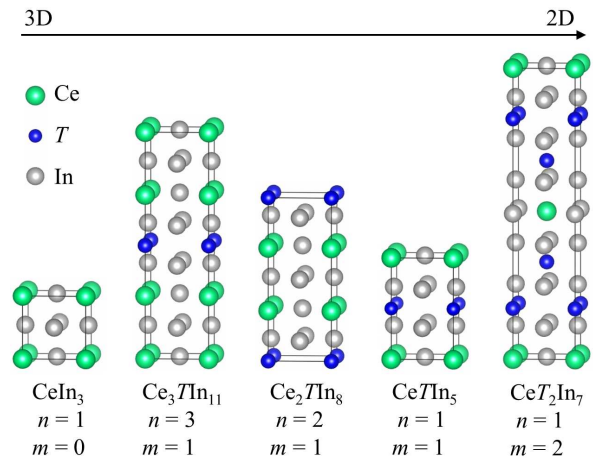}
	\caption{Crystal structure of the Ce$_{n}T_{m}$In$_{3n+2m}$ family, sorted by increasing $m/n$ ratio.}
	\label{fig:family}
\end{figure}

Structurally, CeCoIn$_{5}$ belongs to the family of Ce$_{n}T_{m}$In$_{3n+2m}$ ($T$: transition metal, $n=1$, 2, 3, $m = 0$, 1, 2). Unit cells within this family are composed of $n$ layers of conducting CeIn$_{3}$ separated by $m$ layers of insulating $T$In$_{2}$, as shown in Fig.~\ref{fig:family}. By increasing the ratio $m/n$, the spacing between the CeIn$_{3}$ building blocks becomes larger, which is expected to augment 2D behavior. Therefore, controlling the $m/n$ ratio enables a systematic study of the relation between FS dimensionality and superconducting properties.

Here, we focus on the platinum members of the family, i.e., Ce$_n$Pt$_m$In$_{3n+2m}$. So far, CeIn$_{3}$, Ce$_{3}$PtIn$_{11}$, Ce$_{2}$PtIn$_{8}$, \cite{Kratochvilova_2014} and CePt$_{2}$In$_7$ \cite{Kurenbaeva_2008} have been successfully syn\-the\-sized. Inside this group, CeIn$_{3}$ has the lowest $m/n$ ratio and a cubic crystal structure. Thus, it is expected to host the most three-dimensional (3D) FS sheets, which is supported by both the band-structure calculations~\cite{Settai_2005} and de Haas-van Alphen (dHvA) experiments \cite{Ebihara_1993, Endo_2005}. CeIn$_{3}$ becomes superconducting with $T_c=0.2$~K at a critical pressure $P_c=2.5$~GPa \cite{Mathur_1998}. Increasing the pressure beyond $P_c$ changes the character of the Ce-$4f$ electrons from localized to itinerant \cite{Settai_2005}.

On the other end, CePt$_{2}$In$_7$ has the highest $m/n$ ratio and is, therefore, expected to be the most 2D compound in the Ce$_n$Pt$_m$In$_{3n+2m}$ series. This assumption was ascertained by quantum-oscillation measurements and band-structure calculations \cite{Altarawneh_2011, Miyake_2015, Goetze_2017}. Measurements employing a tunnel-diode-oscillator technique in pulsed magnetic fields gave evidence for the existence of localized $4f$ electrons only above an anomaly at $B_m=45$~T \cite{Altarawneh_2011}. Subsequent experiments clarified that the localized character also persists at lower fields \cite{Miyake_2015, Goetze_2017}. Superconductivity in CePt$_{2}$In$_7$ emerges around a pressure-induced suppression of the AFM ground state at $P_c=\text{3.2-3.5}$~GPa with $T_c=2.1$~K \cite{Bauer_2010, Sidorov_2013, Kurahashi_2015}.

Intermediate $m/n$ ratios are realized in Ce$_{3}$PtIn$_{11}$ and \Ce, which have been synthesized for the first time only recently \cite{Kratochvilova_2014}. Ce$_{3}$PtIn$_{11}$ exhibits AFM order below $T_N=2.2$~K and, unlike CeIn$_{3}$ and CePt$_{2}$In$_7$, becomes superconducting already at ambient pressure with $T_c=0.32$~K, which increases to $T_c=0.7$~K at $P_c=1.3$~GPa\cite{Prokleska_2015}. In addition, its large Sommerfeld coefficient $\gamma_{el}=520$~mJ/mol\,K$^2$ indicates the HF nature of this compound \cite{Prokleska_2015}. Due to the novelty of \Ce, there are, to the best of our knowledge, no publications addressing its ground state so far. However, the isostructural non-magnetic compound Ce$_{2}$PdIn$_{8}$ is known to become superconducting at $T_c=0.7$~K \cite{Kaczorowski_2009, Uhlirova_2010_comm, Kaczorowski_2010_repl, Uhlirova_2010, Kaczorowski_2010}. The same crystal structure is also realized in Ce$_{2}T$In$_{8}$ with $T=$~Co, Rh, Ir.
Two of these compounds are HF superconductors: Ce$_{2}$CoIn$_{8}$ ($T_c=0.4$~K, $\gamma_{el}=1$~J/mol\,K$^2$)\cite{Chen_2002} and Ce$_{2}$RhIn$_{8}$ ($T_c=2$~K at $P_c\approx 0.4$~GPa \cite{Nicklas_2003}, $\gamma_{el}=800$~mJ/mol\,K$^2$ \cite{Ueda_2004}). The third compound, Ce$_{2}$IrIn$_{8}$, is also a HF system ($\gamma_{el}=1.4$~J/mol\,K$^2$). In this material, no superconducting or magnetic order was found down to 50~mK \cite{Thompson_2001}, but a non-Fermi-liquid behavior was observed in applied magnetic fields \cite{Kim_2004}. For all three compounds, quantum-oscillation measurements and/or angle-resolved photoemission spectroscopy revealed quasi-2D FSs~\cite{Ueda_2004, Raj_2005, Souma_2008, Jiang_2015}.

The most direct way to establish the FS topology are quantum-oscillation measurements. In this paper, we report on high-field dHvA measurements of \Ce. Comparing the results to band-structure calculations and to our dHvA data of the $4f$-localized counterpart \Pr\ evidences itinerant Ce-$4f$ electrons in \Ce. Furthermore, our data reveal the existence of a large, 2D FS sheet in this compound. Albeit being less corrugated than its FS counterpart in the isostructural superconductor Ce$_{2}$PdIn$_{8}$, there is no signature of superconductivity in the torque data of Ce$_{2}$PtIn$_{8}$. We also determined the effective masses of Ce$_{2}$PtIn$_{8}$, yielding values between 3.3$m_e$ and 25.7$m_e$, where $m_e$ is the bare electron mass.

\section{Experimental}

\begin{figure}
	\centering
		\includegraphics[width=0.9\columnwidth]{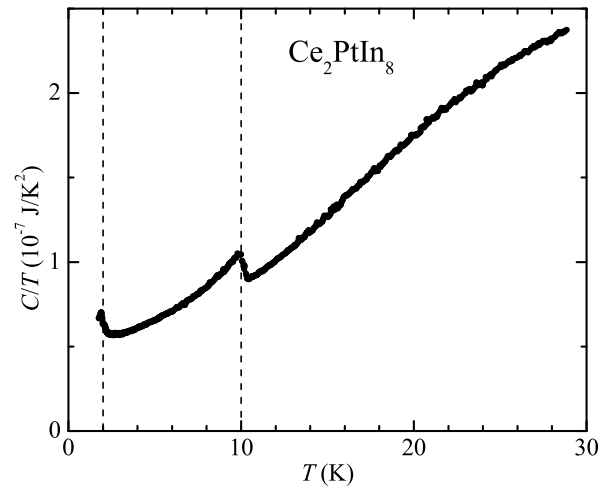}
	\caption{Total heat capacity divided by temperature of the 125~$\mu$g \Ce\ crystal. Dashed lines indicate the known transition temperatures of CeIn$_3$ and Ce$_{3}$PtIn$_{11}$.}
	\label{fig:SpecHt}
\end{figure}

Single crystals of \Ce\ and \Pr\ were grown using the In self-flux method. A detailed description of the growth conditions can be found in Ref.~[\onlinecite{Shishido_2015}]. Both samples used in our study were investigated by Laue diffraction. All the spots we observed in the diffraction images originate from \Ce\ and \Pr, respectively. However, it was previously demonstrated that \Ce\ single crystals grown by In self-flux technique often contain Ce$_{3}$PtIn$_{11}$ and CeIn$_{3}$ impurities, the amount of which depends on the growing conditions~\cite{Kratochvilova_2014}. We have, therefore, measured the specific heat of our \Ce\ sample by relaxation technique in the temperature range from 1.8 to 30~K. The total heat capacity, $C$, divided by temperature, $T$, of the 125~$\mu$g sample is shown in Fig.~\ref{fig:SpecHt}. We have indeed observed two anomalies at 2 and 10~K, which coincide with the AFM transitions in Ce$_{3}$PtIn$_{11}$~\cite{Kratochvilova_2014,Prokleska_2015} and CeIn$_{3}$~\cite{Settai_2005}, respectively. The size of these anomalies, $\Delta C/T \simeq 1.4\times10^{-8}$~J/K$^2$, is small compared to the large specific-heat jumps at $T_N$ in pure Ce$_{3}$PtIn$_{11}$ ($\Delta C/T = 2.4$~J/mol\,K$^2$)~\cite{Kratochvilova_2014,Prokleska_2015} and CeIn$_{3}$ ($\Delta C/T = 1.1$~J/mol\,K$^2$)~\cite{Settai_2005}. We, thus, estimate that our \Ce\ sample contains a mass fraction of approximately 8~\%\ Ce$_{3}$PtIn$_{11}$ and 5~\%\ CeIn$_{3}$ impurities. This is taken into account during the data analysis, and does not affect our main conclusions. After subtracting impurity contributions, we estimated a Sommerfeld coefficient $\gamma=0.5$~J/mol\,K$^2$.

Angle-dependent quantum-oscillation measurements were performed on the same sample using capacitive torque magnetometry, employing 25~$\mu$m and 50~$\mu$m thick CuBe cantilevers. The experiments were conducted in a dilution refrigerator with a base temperature of about 30~mK. High magnetic fields up to 34~T were provided by a resistive magnet in the LNCMI-Grenoble.

\section{Results and Discussion}

\subsection{dHvA measurements of \Ce}

\begin{figure}
	\centering
		\includegraphics[width=0.87\columnwidth]{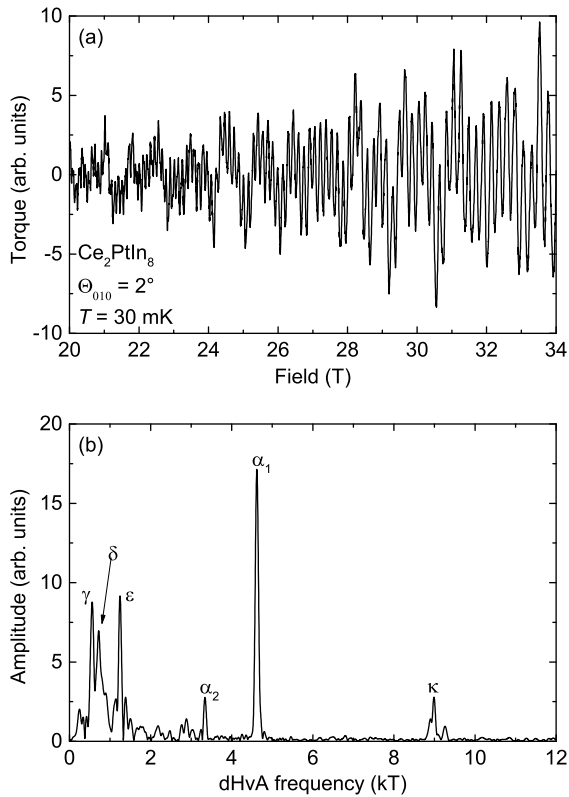}
	\caption{(a) Torque signal of \Ce\ after subtracting a non-oscillatory background signal. (b) Corresponding frequency spectrum obtained by a Fourier transform of the data shown in (a).}
    \label{fig:CePtIn_example}
\end{figure}

Figure \ref{fig:CePtIn_example}(a) shows a typical torque signal for \Ce\ taken at an angle $\Theta_{010}=2^\circ$ after subtracting the non-oscillatory background. Throughout this paper, all angles are measured from the crystallographic $c$ axis. The quantum oscillations are clearly visible. By performing a Fourier transformation, we obtained the corresponding frequency spectrum shown in Fig. \ref{fig:CePtIn_example}(b). There are six fundamental frequencies, denoted as $\alpha_{1}$, $\alpha_{2}$, $\gamma$, $\delta$, $\varepsilon$, and $\kappa$. Details of the angular dependence of the dHvA frequencies will be discussed in Sec.~\ref{sec:exp_vs_calc}, together with the results of the band-structure calculations. No signatures of phase transitions were observed in any of the torque signals.

\subsection{Band-structure calculations}

\begin{figure}
	\centering
		\includegraphics[width=0.9\columnwidth]{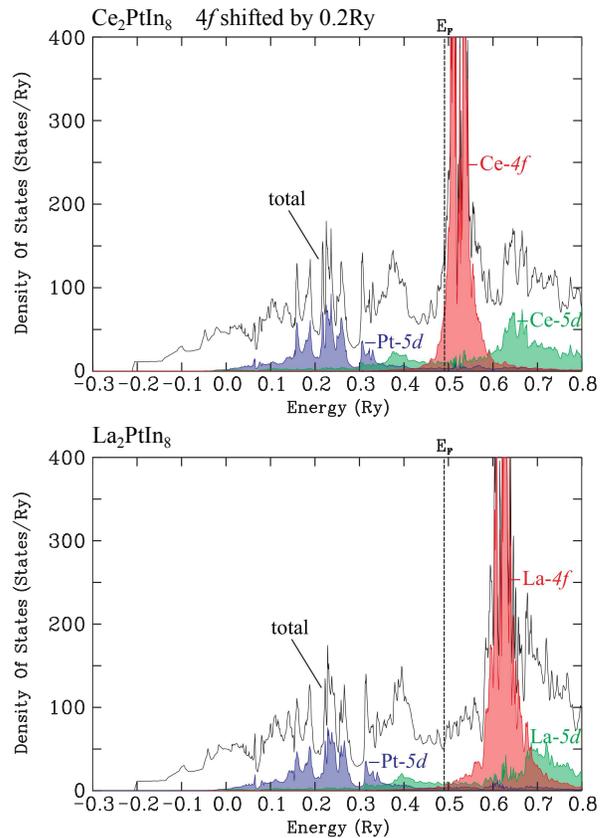}
	\caption{Calculated densities of states for \Ce\ (top) and \La\ (bottom). Dashed lines indicate the Fermi energy $E_F$.}
	\label{fig:DOS}
\end{figure}

\begin{figure*}
	\centering
		\includegraphics[width=0.9\textwidth]{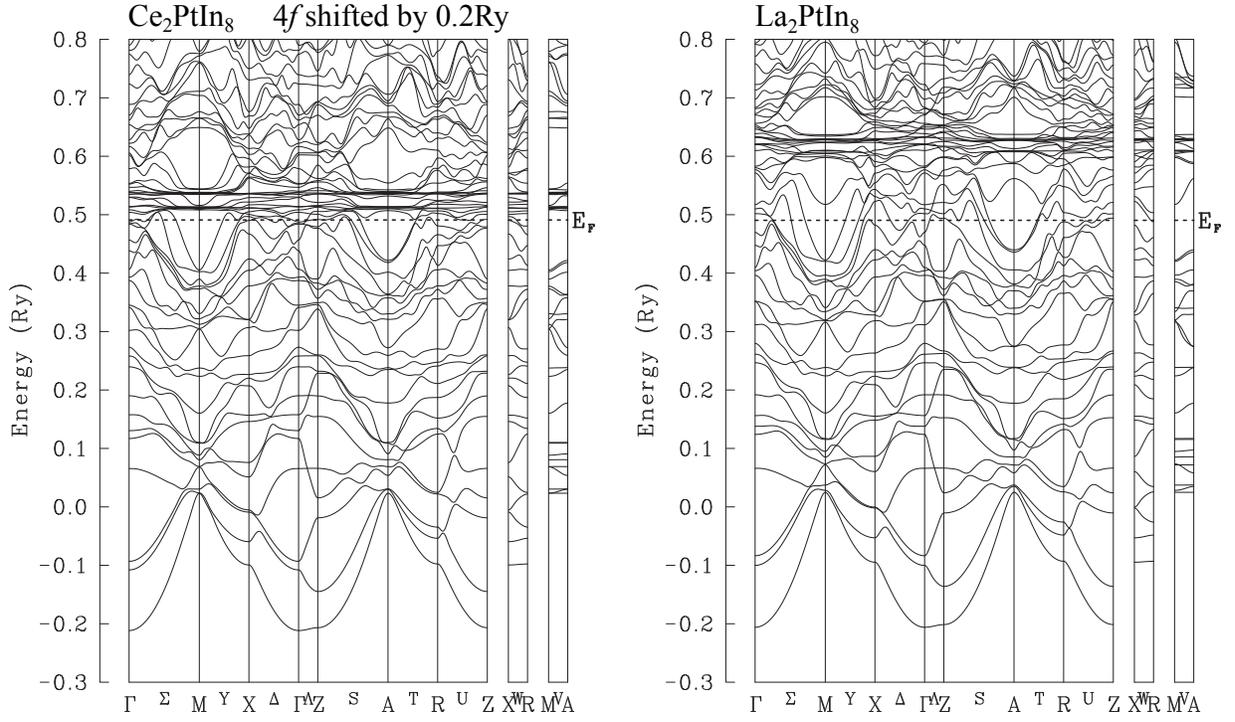}
	\caption{Calculated band structures along high-symmetry axes for \Ce\ (left) and \La\ (right). Dashed lines indicate the Fermi energy $E_F$.}
	\label{fig:band_structure}
\end{figure*}

\begin{figure*}
	\centering
		\includegraphics[width=0.95\textwidth]{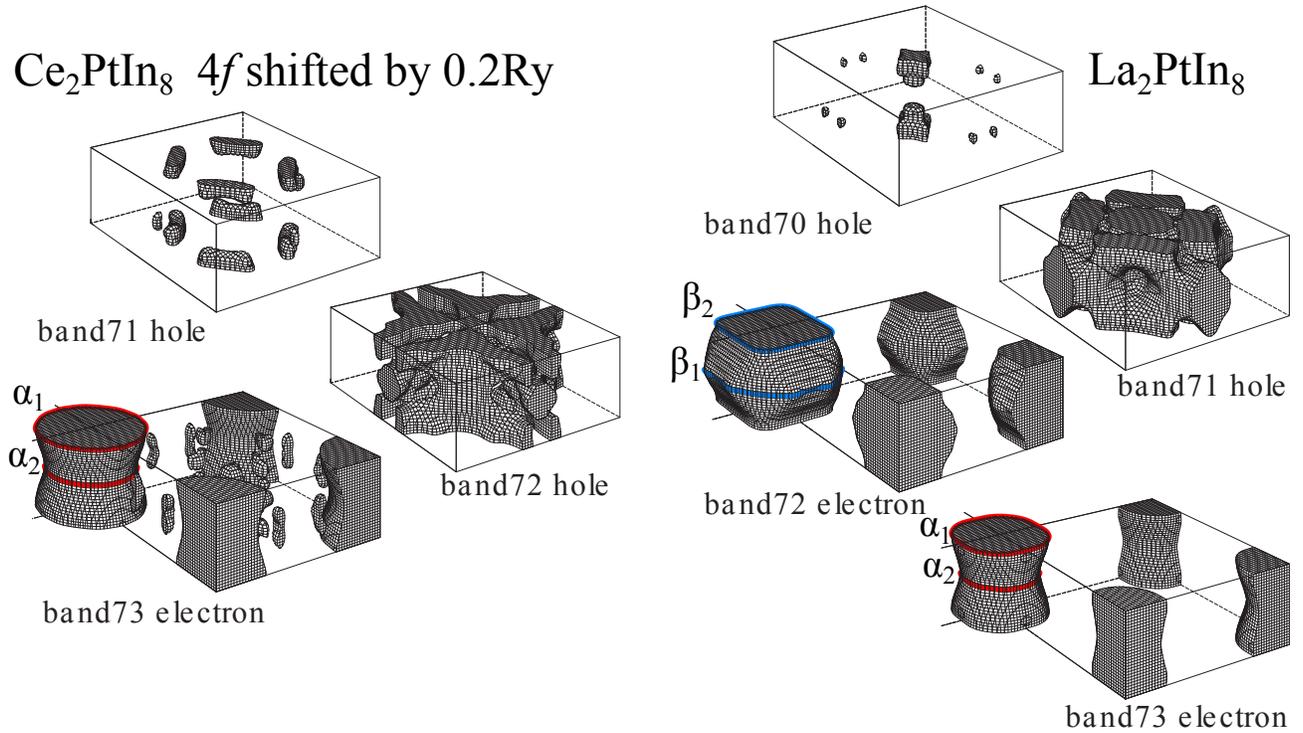}
	\caption{Calculated FSs for \Ce\ (left) and \La\ (right). Solid lines indicate the extremal cross-sections of quasi-2D FS sheets for fields applied parallel to the $c$ axis.}
	\label{fig:CePtIn_FS_joint}
\end{figure*}

In order to determine whether the $4f$ electrons are itinerant or localized, we performed band-structure calculations using the \textsc{kansai} code, based on an FLAPW (full potential linear augmented plane wave) method with a local density approximation (LDA). The relativistic effect is considered by using the technique proposed by Koelling and Harmon~\cite{Koelling_1977}. The spin-orbit coupling is included as the second variational procedure for valence electrons.

In the calculations, we used the lattice parameters of $a=4.6893$~\AA\ and $c=12.1490$~\AA, taken from Ref.~[\onlinecite{Kratochvilova_2014}]. Structure parameters are 0.30795 for the $2d$ site of Ce, 0.12241 for the $4i$ site of In, and 0.3072 for the $2h$ site of In. Core electrons (Xe core minus $5s^{2}5p^6$ electrons for Ce/La, Xe core minus $5p^6$ plus $4f^{14}$ for Pt, Kr core for In) are calculated inside the Muffin-Tin spheres in each self-consistent step. $5s^{2}5p^6$ electrons on Ce/La, $5p^6$ electrons on Pt, and $4d^{10}$ electrons on In are calculated as valence electrons by using a second energy window. The LAPW basis functions are truncated at $|k+G_i| \le 4.20 \times 2\pi/a$, corresponding to 821 LAPW functions. The sampling points are 220 $k$-points uniformly distributed in the irreducible 1/16th of the Brillouin zone, which are divided by (18, 18, 6).

In order to improve the agreement between the calculations and our Ce$_2$PtIn$_8$ dHvA data, the Ce-$4f$ states are shifted upward by 0.2~Ry from LDA. To compare the non-$f$ reference or localized $4f$ electrons model, the calculation for La$_2$PtIn$_8$ was also performed by using the structural parameters of Ce$_2$PtIn$_8$ as crystals of La$_2$PtIn$_8$ are currently unavailable. For both itinerant and localized $4f$ electrons, the resulting densities of states and band structures are shown in Figs.~\ref{fig:DOS} and \ref{fig:band_structure}, respectively. The corresponding Fermi surfaces are shown in Fig.~\ref{fig:CePtIn_FS_joint}.

\subsection{Experimental data of \Ce\ vs.~band-structure calculations}
\label{sec:exp_vs_calc}

\begin{figure*}
	\centering
		\includegraphics[width=0.95\textwidth]{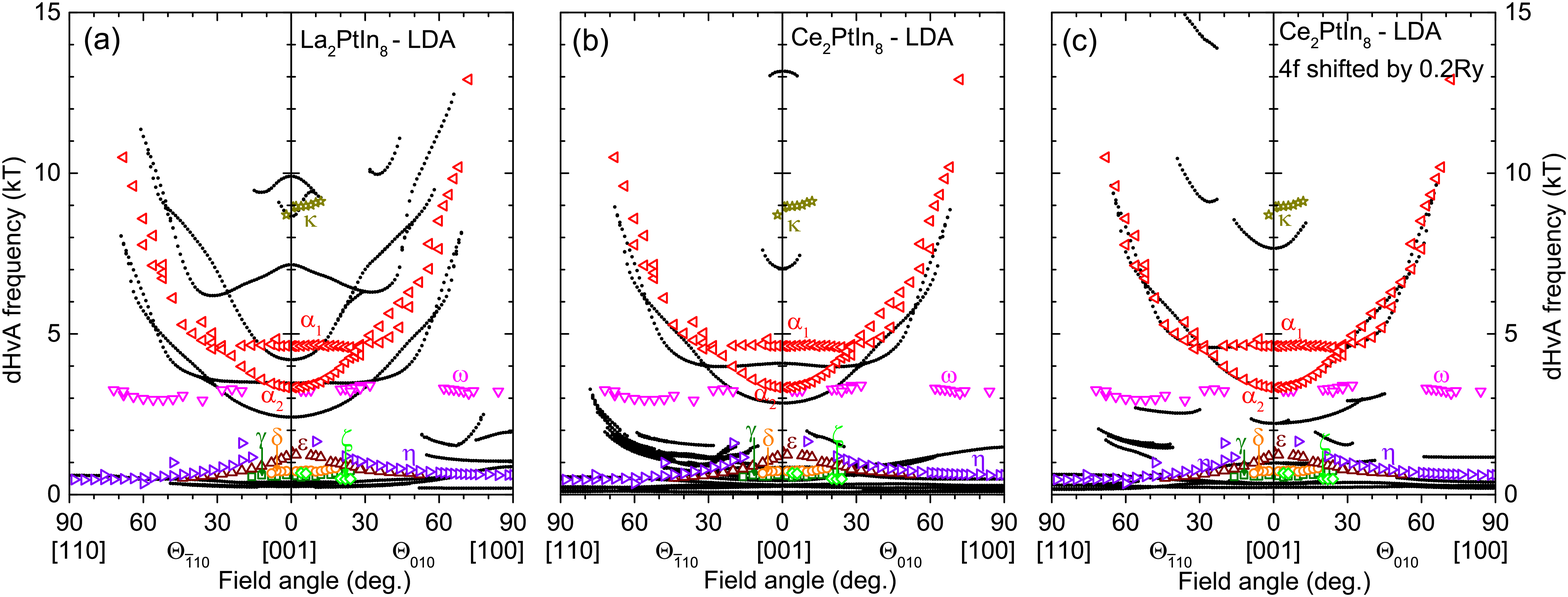}
	\caption{Comparison of experimental dHvA frequencies of \Ce\ (open symbols) with calculated extremal cross sections (solid circles) of (a) \La, (b) \Ce, and (c) \Ce\ with shifted $4f$ states. Calculated dHvA frequencies lower than 200 T are not shown for clarity.}
    \label{fig:CePtIn_exp_vs_calc}
\end{figure*}

The comparison of the experimentally determined dHvA frequencies in \Ce\ and the calculated frequencies for \La\ and \Ce\ is shown in Fig.~\ref{fig:CePtIn_exp_vs_calc}. Overall, we observed eight dHvA frequency branches, denoted by $\alpha$, $\gamma$, $\delta$, $\varepsilon$, $\zeta$, $\eta$, $\kappa$, and $\omega$. One of these branches, $\alpha$, could be traced up to 70$^\circ$ away from the $c$ axis. Roughly following a $1/\cos(\Theta)$ behavior, it indicates the existence of a quasi-2D corrugated cylindrical FS sheet. In addition, a multitude of frequencies lower than 2~kT suggests the existence of several small FS pockets. Two of these frequencies, $\varepsilon$ and $\eta$, were observed over the entire angular range. Surprisingly, we also observed a fairly large and almost spherical FS sheet, evidenced by the branch $\omega$ with the dHvA frequency of about 3~kT. Considering the tetragonal crystal structure with a $c/a$ ratio of about 2.6, near-spherical FS pockets of this size are unexpected. The highest frequency, $\kappa$, appears only close to the $c$ axis. It was observed only for the rotation from [001] to [100] direction, because we used a thinner and, therefore, more sensitive cantilever for small angles within this plane.

Due to the layered structure of \La\ and \Ce, both the calculations for localized and for itinerant $4f$ electrons suggest the existence of quasi-2D FSs. The main difference, however, is that the calculation for localized $4f$ electrons predicts the existence of two quasi-2D FSs, opposed to only one in the calculation for itinerant $4f$ electrons (see Fig.~\ref{fig:CePtIn_FS_joint}). Since 2D FSs feature a favorable curvature factor and a $1/\cos(\Theta)$ angular dependence of their extremal cross section, they are easily detectable by torque magnetometry. Therefore, the experimental observation of only one 2D FS sheet with extremal areas $\alpha_1$ and $\alpha_2$ strongly supports the calculation for itinerant $4f$ electrons. Moreover, the direct comparison of the experimental data with the \La\ calculations, shown in Fig.~\ref{fig:CePtIn_exp_vs_calc}(a), yields a very poor agreement. In particular, none of the observed smaller dHvA frequencies $\gamma \ldots \eta$ can be attributed to a calculated branch, especially near the $c$ axis. For $\alpha$, there is a calculated branch exhibiting a similar angular dependence, but the frequencies differ by about 1~kT. 

In contrast, most of the high- and low-frequency experimental branches are reproduced by the \Ce\ calculations, as shown in Fig.~\ref{fig:CePtIn_exp_vs_calc}(b). All the low frequencies $\gamma \ldots \eta$ are in good qualitative agreement with calculated ones. More importantly, the frequency associated with the almost cylindrical FS sheet, $\alpha$, is in very good qualitative agreement with the corresponding calculated branch. We could further improve the quantitative agreement of the quasi-2D FS sheet by shifting the $4f$ states by $0.2$~Ry, as depicted in Fig.~\ref{fig:CePtIn_exp_vs_calc}(c). However, there are also a few mismatches. Several calculated branches originating from band 72 were not observed experimentally. Presumably, this is due to an unfavorable curvature factor. We note, that the calculated frequency branches from band 72 were not observed in the isostructural compound Ce$_{2}$PdIn$_{8}$ as well \cite{Goetze_2015}.

The frequencies $\omega$ and $\kappa$ are not explained by theory. Assuming an epitaxial growth of the CeIn$_{3}$ impurities in our crystal, $\omega$ coincides with the frequency branch $d$ of CeIn$_{3}$ \cite{Ebihara_1993}, rendering CeIn$_{3}$ impurities to be the most likely source of $\omega$. On the other hand, the origin of $\kappa$ is yet unknown. Since the angular dependence of this branch does not reflect the cubic symmetry of CeIn$_{3}$ and does not coincide with any of the known frequency branches of CeIn$_{3}$, it appears unlikely for $\kappa$ to originate from the CeIn$_3$ impurity. However, the Ce$_{3}$PtIn$_{11}$ impurity phase, also present in our sample, cannot be ruled out as a possible source of the branch $\kappa$. Quantum-oscillation or ARPES measurements of Ce$_{3}$PtIn$_{11}$ are required to verify this hypothesis. Apart from the aforementioned branches, the band-structure calculations for \Ce\ with shifted $4f$-electron bands yield an excellent agreement with the experimental data, strongly suggesting itinerant Ce-$4f$ electrons in this compound.

\subsection{Comparison with \Pr}

\begin{figure}
	\centering
		\includegraphics[width=0.87\columnwidth]{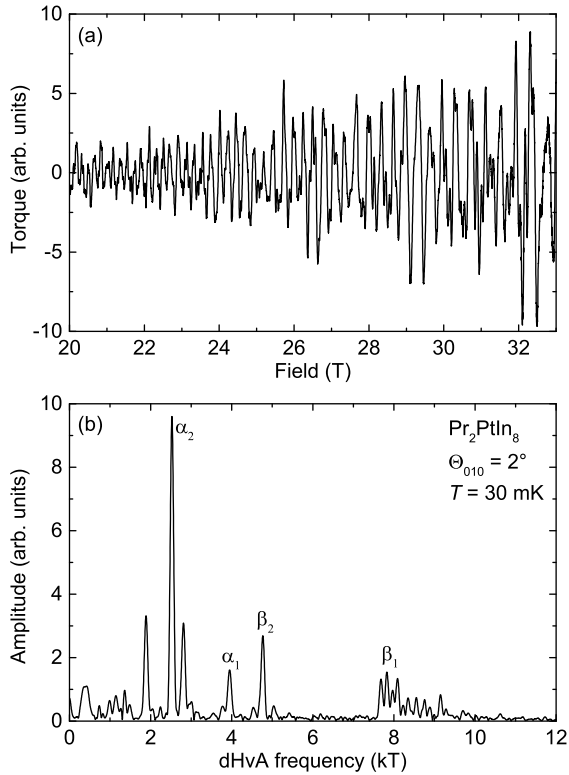}
	\caption{(a) Torque signal of \Pr\ after subtracting a non-oscillatory background, taken at the same angle as the signal for \Ce\ shown in Fig.~\ref{fig:CePtIn_example}. (b) Corresponding frequency spectrum obtained by a Fourier transform of the data shown in (a).}
    \label{fig:PrPtIn_example}
\end{figure}

\begin{figure}
	\centering
		\includegraphics[width=0.97\columnwidth]{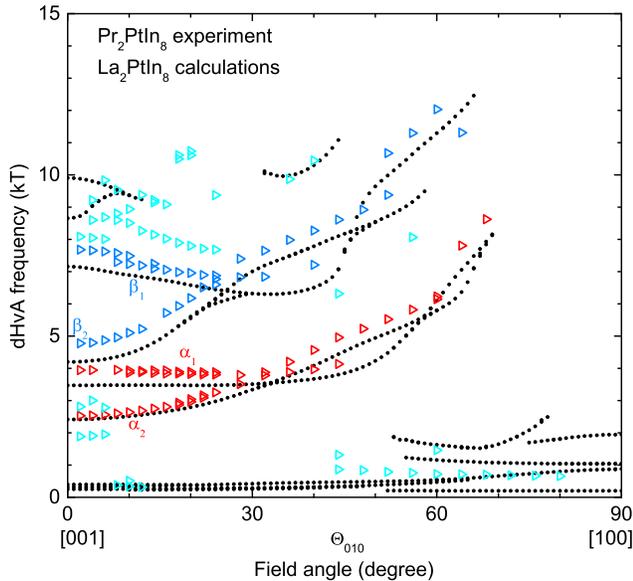}
	\caption{Comparison of the experimental dHvA frequencies of \Pr\ (open symbols) with calculated extremal cross sections (solid circles) of \La.}
	\label{fig:PrPtIn_vs_LaPtIn}
\end{figure}

\begin{figure}
	\centering
		\includegraphics[width=0.97\columnwidth]{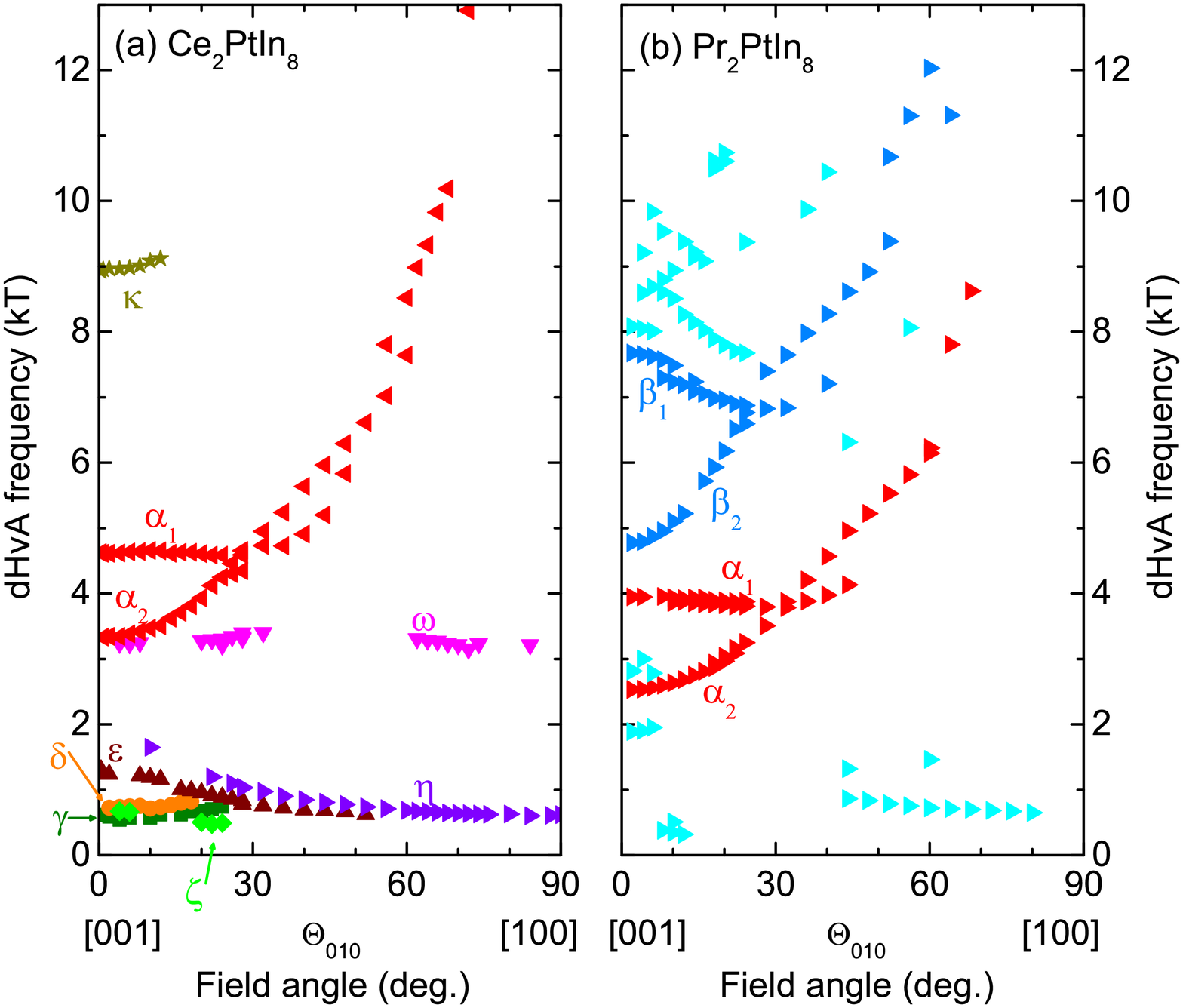}
	\caption{Angular dependence of the experimental dHvA frequencies in (a) \Ce\ and (b) \Pr.}
	\label{fig:PrPtIn_vs_CePtIn}
\end{figure}

It would be desirable to directly compare the dHvA oscillations in \Ce\ with those measured in \La\ in order to establish the localized or itinerant character of the Ce-$4f$ electrons beyond any doubts. However, single crystals of \La\ are currently unavailable, rendering a direct comparison impossible. Since the Pr-based analogues of Ce-based HF compounds are known to have localized $4f$ electrons, e.g.~in PrCoIn$_{5}$ \cite{Elgazzar_2008}, \Pr\ poses an alternative candidate for comparison. We, therefore, performed dHvA measurements on \Pr. In addition, we conducted specific-heat measurements on our \Pr\ single crystal, revealing a Sommerfeld coefficient of $\gamma=90$~mJ/mol\,K$^2$. This evidences greatly reduced electronic correlations in \Pr\ compared to \Ce, as expected for localized $4f$ electrons.

Figure \ref{fig:PrPtIn_example}(a) shows a typical torque signal for \Pr, after subtracting the non-oscillatory background, taken at the same angle, $\Theta_{010}=2^\circ$, as shown in Fig.~\ref{fig:CePtIn_example} for \Ce. The corresponding dHvA frequency spectrum in Fig.~\ref{fig:PrPtIn_example}(b) is clearly different from the one in Fig.~\ref{fig:CePtIn_example}(b), suggesting different topologies of the FSs of \Ce\ and \Pr. In order to verify the consistency, Fig.~\ref{fig:PrPtIn_vs_LaPtIn} compares calculated frequencies for $4f$-localized \La\ with the experimental data of \Pr. Indeed, there is a very good agreement between the two data sets. All the large FS sheets calculated for bands 71, 72, and 73 are qualitatively supported by the experimental data. Thus, both the results of the \La\ band-structure calculations and the \Pr\ dHvA measurements are a good reference for $4f$-localized \Ce.

When juxtaposing the angular dependences of \Ce\ and \Pr, clear FS differences become even more obvious, as shown in Fig.~\ref{fig:PrPtIn_vs_CePtIn}. Apart from parts of the branch $\eta$, none of the branches of \Ce\ can be found in \Pr. Consequently, the electronic states in \Ce\ differ from those in the $4f$-localized \Pr, further proving the itinerant character of the Ce-$4f$ electrons in \Ce. Although \Ce\ is an intermediate compound between CeIn$_{3}$ and CePt$_{2}$In$_{7}$ from a structural point of view, it does not share their localized Ce-$4f$ character.

\subsection{Effective masses of \Ce}

\begin{figure}
	\centering
		\includegraphics[width=0.97\columnwidth]{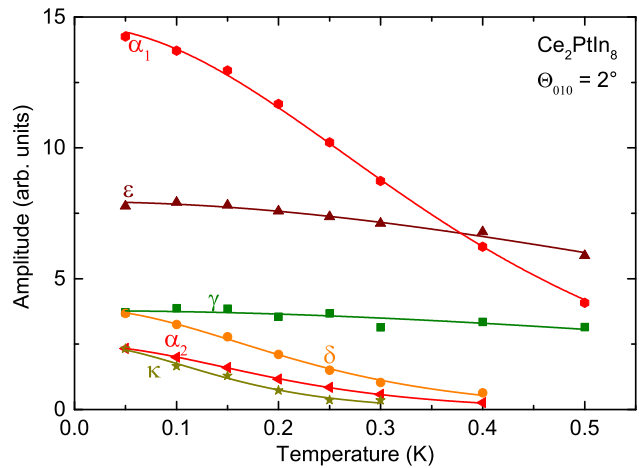}
	\caption{Temperature-dependent amplitudes, taken at the same angle as the data shown in Fig.~\ref{fig:CePtIn_example}, with fit lines by use of the Lifshitz-Kosevich formula~\cite{Shoenberg_1984}.}
    \label{fig:CePtIn_temp_dep}
\end{figure}

We determined the effective masses of \Ce\ from the temperature dependence of the dHvA oscillation amplitudes. These amplitudes were measured in the field range of 20 to 34~T, for temperatures between 50 and 500~mK, at the same angle as the data shown in Fig.~\ref{fig:CePtIn_example}. According to the Lifshitz-Kosevich formula, this dependence is proportional to $x/\sinh x$, where $x=\alpha T m^\ast /B$ and $\alpha=14.69$~T/K~\cite{Shoenberg_1984}. Here, $m^\ast$ represents the effective mass given in multiples of the bare electron mass $m_e$. Figure \ref{fig:CePtIn_temp_dep} shows the fit of the temperature dependence of the oscillatory amplitudes by this formula, yielding $m^\ast$ as fit parameter. The results are summarized in Table~\ref{tab:CePtIn_masses}, where the band masses, $m_b$, obtained from band-structure calculations with shifted $4f$ states by taking the derivative $dF/dE$, are also shown.

The effective masses are fairly large, ranging from 3.3$m_e$ for $\gamma$ to 25.7$m_e$ for $\kappa$. The higher the mass of a specific orbit, the stronger the damping of the corresponding oscillation by impurity scattering \cite{Shoenberg_1984}. Given the strongly enhanced effective mass of the $\kappa$ branch, it seems unlikely that it originates from the Ce$_{3}$PtIn$_{11}$ impurity, because it would require a very high crystal quality of the impurity itself. Similarly, we observed the impurity frequency $\omega$ only for field directions where its effective mass is as small as $2m_e$ \cite{Ebihara_1993}.

The measured effective masses exceed the calculated band masses by about an order of magnitude or more. This implies that there is a significant mass enhancement due to many-body interactions, which are not included in the LDA calculations. Effective masses much higher than calculated were observed in other non-magnetic HF compounds with itinerant $4f$ electrons, such as CeCoIn$_5$~\cite{Settai_2001} and Ce$_2$PdIn$_8$~\cite{Goetze_2015}. For the $\alpha_1$ and $\alpha_2$ orbits, both lying on the same FS sheet, the mass-enhancement factors are 9.5 and 30, respectively. Such largely different mass enhancements on a single FS sheet were also found in CeCoIn$_5$~\cite{Settai_2001} and CeIrIn$_5$~\cite{Haga_2001}.  In CeCoIn$_5$, the effective masses were reported to strongly decrease with magnetic field~\cite{Settai_2001}. Similarly, the Sommerfeld coefficient of the specific heat in Ce$_2$PdIn$_8$ is strongly suppressed by magnetic field~\cite{Tokiwa_2011}. This is due to a close proximity to a QCP of these two compounds. In \Ce, on the contrary, we have not observed any appreciable field dependence of the effective masses over the field range 16\---34 T of our measurements. This suggests that \Ce\ is located further away from a QCP.

In comparison to AFM CeIn$_{3}$ ($m^{\ast}=2 \ldots 12$$m_e$ \cite{Umehara_1990, Ebihara_1993}) and CePt$_{2}$In$_{7}$ ($m^{\ast}=1.3 \ldots 6.2$$m_e$ \cite{Altarawneh_2011, Goetze_2017}) with localized $4f$ electrons, the effective masses in \Ce\ are significantly higher. Again, this reflects the itinerant character of the Ce-$4f$ electrons, leading to an increased density of states at the Fermi level and stronger many-body interactions.

\begin{table}
	\centering
		\begin{tabular*}{\columnwidth}{@{\extracolsep{\stretch{1}}}*{5}{ccccc}@{}}
			\hline
			\hline
                  & \multicolumn{2}{c}{Experiment}  & \multicolumn{2}{c}{Calculation}											\\ 
           Branch & $F$ (kT) 	& \multicolumn{1}{c}{$m^{\ast}$ ($m_{e}$)}	& $F$ (kT)& $m_{b}$ ($m_{e}$)	\\ 
			\hline
										&						&											& 0.227		&	0.67		\\ 
										&						&											& 0.337		&	0.35		\\ 
										&						&											& 0.355		&	0.25		\\ 
			$\gamma$			&	0.55			&	3.3(1)							&					&					\\ 
			$\delta$			&	0.73			&	15.0(3)							&					&					\\ 
										&						&											& 0.983		&	1.46		\\ 
			$\varepsilon$	&	1.24			&	4.6(1)							&					&					\\ 
										&						&											& 2.222		&	2.14		\\ 
			$\alpha_{2}$	&	3.34			&	19.2(1)							&	3.357		&	0.61		\\ 
			$\alpha_{1}$	&	4.62			&	10.5(2)							&	4.719		&	1.00		\\ 
										&						&											& 7.676		&	1.64		\\ 
			$\kappa$			&	8.99			&	25.7(10)						&					&					\\ 
			\hline
			\hline
		\end{tabular*}
		\caption{Experimental and calculated (with shifted 4$f$ states) dHvA frequencies and effective masses of \Ce\ for $\Theta_{010}=2^\circ$. Branch assignments refer to Fig.~\ref{fig:CePtIn_exp_vs_calc}.}
\label{tab:CePtIn_masses}
\end{table}

\subsection{Fermi-surface dimensionality}

In order to quantify the degree of deviation from ideal two-dimensionality, we previously introduced the value $\Delta=(S_{max}-S_{min})/S_{avg}$ \cite{Goetze_2015}. Here, $S_{min}$, $S_{avg}$, and $S_{max}$ are the minimum, average, and maximum cross sections of a quasi-2D FS sheet, respectively. Due to the fact that the observed dHvA frequencies are proportional to the extremal cross sections, we can calculate $\Delta$ from our experimental data.

As shown in Fig.~\ref{fig:CePtIn_FS_joint}, one of the calculated FS sheets of \Ce\ is nearly cylindrical. The observation of the experimental branch $\alpha$ proves the existence of this 2D sheet [see Fig.~\ref{fig:CePtIn_exp_vs_calc}(b)]. For this FS sheet, we find a value of $\Delta=0.32$. This compares to 0.04, 0.03, and 0.07 for the three cylindrical sheets of CePt$_{2}$In$_7$ \cite{Goetze_2017}. Also in the prototypical compound, CeCoIn$_{5}$, the value $\Delta=0.21$ (averaged from Refs.~[\onlinecite{Settai_2001}], [\onlinecite{Hall_2001}], and [\onlinecite{Polyakov_2012}]) is smaller. Consequently, the FS dimensionality of \Ce\ lies between those of CeIn$_{3}$ and CeCoIn$_{5}$, as expected from the intermediate two-dimensionality of the crystal structure.

\section{Conclusions}

\begin{figure}
	\centering
		\includegraphics[width=0.97\columnwidth]{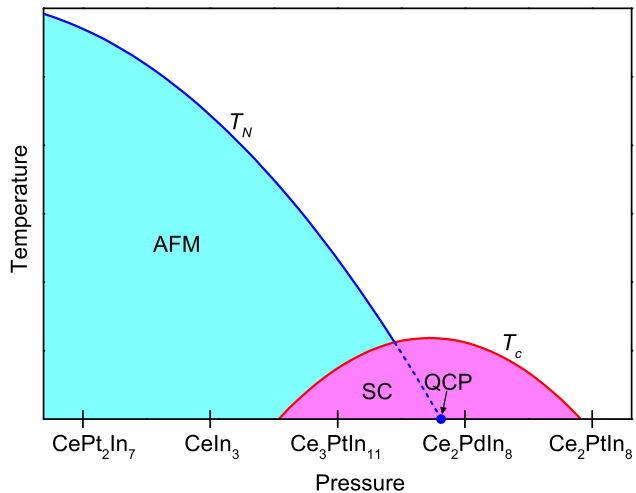}
	\caption{Schematic location of \Ce\ and other Ce$_{n}T_{m}$In$_{3n+2m}$ compounds in the generic phase diagram of AFM QCPs in Ce-based HF systems. Note that CePt$_{2}$In$_{7}$ and CeIn$_{3}$ were arranged according to their critical pressures ($\approx 3.3$ and 2.5~GPa, respectively), but not according to their N\'{e}el temperatures (5.5 and 10~K, respectively)\cite{Mathur_1998, Bauer_2010, Sidorov_2013, Kurahashi_2015}. \Ce\ and Ce$_{2}$PdIn$_{8}$ were arranged according to their unit cell volumes\cite{Kratochvilova_2014}.}
    \label{fig:phase_diag}
\end{figure}

In summary, we conducted measurements of the dHvA effect on the HF compound \Ce\ and its $4f$-localized counterpart \Pr. By contrasting the data of these two compounds and by comparing the \Ce\ data to band-structure calculations, we find clear evidence for itinerant Ce-$4f$ electrons in this material. In addition, the effective masses are strongly enhanced, varying between 3.3$m_e$ and 25.7$m_e$ for the different FS sheets.

Structurally, \Ce\ can be considered as an intermediate compound between CeIn$_{3}$ and CePt$_{2}$In$_7$.
The FSs of \Ce\ follow the structural trend, being less 2D than those of CePt$_{2}$In$_7$, but more than those of CeIn$_{3}$.
Both CeIn$_{3}$ and CePt$_{2}$In$_7$ have localized Ce-$4f$ electrons, in contrast to \Ce.
It is, therefore, not surprising that the effective masses in \Ce\ are considerably enhanced.

There are a lot of similarities between the electronic structures of \Ce\ and the isostructural Ce$_{2}$PdIn$_{8}$ \cite{Goetze_2015}.
All the FS sheets of \Ce\ correspond to almost identical sheets in Ce$_{2}$PdIn$_{8}$.
The only exception are very small FS pockets originating from band 74 of Ce$_{2}$PdIn$_{8}$ \cite{Goetze_2015}, which cannot be found in \Ce.
Both compounds feature enhanced effective masses and itinerant $4f$ electrons.

In the generic temperature-pressure phase diagram of Ce-based HF compounds, the AFM ordering temperature $T_N$ decreases upon increasing pressure \cite{Flouquet_2011}, as shown in Fig.~\ref{fig:phase_diag}.
At $T=0$ and the critical pressure $P_c$, a QCP separates the AFM and the paramagnetic phases.
The $f$ electrons of Ce are usually found to be localized in the former, and itinerant in the latter~\cite{Araki2001, Shishido2005, Settai_2005}.
The superconducting dome is located around this critical pressure. In Ce$_{2}$PdIn$_{8}$, for example, the observation of a non-Fermi-liquid behavior at low temperature~\cite{Dong2011,Tran2011,Gnida2012,Tokiwa_2011,Matusiak2011,Fukazawa2012,Tran2012} suggests the close proximity to a QCP.

Our data reveal the $4f$-itinerant ground state of \Ce, and no signatures of a superconducting transition.
This is not surprising, given that \Ce\ has a smaller unit cell volume than Ce$_{2}$PdIn$_{8}$\cite{Kratochvilova_2014}, but very similar FSs.
Consequently, \Ce\ can be considered a compressed version of Ce$_{2}$PdIn$_{8}$.
Altogether, this suggests that in the temperature-pressure phase diagram, \Ce\ is located outside the superconducting dome at pressures larger than $P_c$ (see Fig.~\ref{fig:phase_diag}).
Thus, we assume that a negative pressure induced by chemical substitution would tune \Ce\ to an AFM QCP, similar to CeCu$_{6-x}$Au$_{x}$ \cite{Germann_1988, Lees_1988, Germann_1989} or Ce$_{1-x}$La$_{x}$Ru$_{2}$Si$_{2}$ \cite{Besnus_1987}.

In Ref.~[\onlinecite{Goetze_2015}], we found a trend between the FS dimensionality and the superconducting $T_c$ of several Ce-based HF compounds.
The corrugated cylindrical FS sheet of \Ce\ ($\Delta=0.32$) deviates less from ideal two-dimensionality than the corresponding sheet in Ce$_{2}$PdIn$_{8}$ ($\Delta=0.39$) \cite{Goetze_2015}.
Therefore, one could expect to observe superconductivity in \Ce\ with a higher $T_c$ than in Ce$_{2}$PdIn$_{8}$.
However, we have not observed any sign of superconductivity in \Ce\ in our torque measurements.
Since \Ce\ is likely to be located further away from a QCP, we conclude that the proximity to a QCP is essential for superconductivity.
The other known members of the Ce$_{n}$Pt$_{m}$In$_{3n+2m}$ family can be tuned to a QCP by applying pressure.
With increasing structural two-dimensionality, their respective $T_c$'s at the QCP increase from 0.2~K (CeIn$_3$ \cite{Mathur_1998}) to 0.7~K (Ce$_3$PtIn$_{11}$\cite{Prokleska_2015}) to 2.1~K (CePt$_2$In$_{7}$ \cite{Bauer_2010, Sidorov_2013, Kurahashi_2015}).
For CeIn$_3$~\cite{Settai_2005, Ebihara_1993, Endo_2005}, \Ce\ (this work), and CePt$_2$In$_{7}$\cite{Altarawneh_2011, Miyake_2015, Goetze_2017}, the FS dimensionality follows the structural dimensionality.
Assuming that the FSs of Ce$_3$PtIn$_{11}$ also follow this trend, this indicates that quasi-2D FSs may enhance the $T_c$ observed at the QCP.
An investigation of the Ce$_3$PtIn$_{11}$ FSs would be desirable to verify this trend.

\section{Acknowledgments}

We acknowledge the support of the LNCMI-CNRS, and the HLD-HZDR, members of the European Magnetic Field Laboratory (EMFL), ANR-DFG grant ``Fermi-NESt'', JSPS KAKENHI Grants No. JP15H05882, No. JP15H05884, No. JP15H05886, No. JP15K21732 (J-Physics), and JP16H02450. K.G. acknowledges support from the DFG within GRK 1621.

\bibliography{Ce2PtIn8_bibliography}

\end{document}